\documentstyle[epsf.v1.13]{elsart}
\newcommand{\lsim}{\mathrel{\rlap{\lower4pt\hbox{\hskip1pt$\sim$}}
\raise1pt\hbox{$<$}}}
\newcommand{\sfrac}[2]{\mbox{\footnotesize $\frac{#1}{#2}$}}
\begin{document}
\begin{frontmatter}
%_______________________ Title, Authors ____________________________________
%{\hspace*{\fill}\parbox{55mm}
{\sc ANL-PHY-8302-TH-96}\hspace*{\fill}nucl-th/9602012

\title{Goldstone Theorem and Diquark Confinement\\Beyond Rainbow-Ladder
Approximation} 
\author{A. Bender, C. D. Roberts and L. v. Smekal}
\address{Physics Division, Bldg. 203, \\
Argonne National Laboratory, Argonne IL 60439-4843}
%-------------------------------------------------------------------
\begin{abstract}
The quark Dyson-Schwinger equation and meson Bethe-Salpeter equation are
studied in a truncation scheme that extends the rainbow-ladder approximation
such that, in the chiral limit, the isovector, pseudoscalar meson remains
massless.  Quark-quark (diquark) correlations, which are bound in
rainbow-ladder approximation, are destabilised by repulsive contributions
that only appear at higher order in the Bethe-Salpeter kernel.  The net
effect of higher order terms on the meson bound-state masses is small.\\
\end{abstract}
\begin{keyword}
Hadron spectroscopy; Confinement;
Dyson-Schwinger equations; Bethe-Salpeter equation; 
Nonperturbative QCD phenomenology. \\[6pt]
PACS Numbers: 12.38.Aw, 11.30.Rd, 24.85.+p, 12.38.Lg
\end{keyword}
\end{frontmatter}
%____________________________________________________________

{\bf 1. Introduction}.\\ 
%\section{Introduction}
The analytic structure of quark and gluon propagators can provide an
understanding of confinement.  Should these propagators have no Lehmann
representation then a sufficient condition for confinement is satisfied;
i.e., the absence of quark and gluon production thresholds in ${\cal
S}$-matrix elements describing colour-singlet to singlet transitions.  This
in itself, however, neither precludes nor entails the existence of
bound-states, whether coloured or not.  The existence of two-body
bound-states can only be established via the Bethe-Salpeter equation [BSE].

The homogeneous Bethe-Salpeter equation is derived under the assumption that
the associated two-body ${\cal T}$-matrix has a pole in a given channel.  The
absence of a solution to the equation contradicts this assumption and
establishes that no bound-state exists with the quantum numbers of the
channel under consideration.

The intrinsically nonperturbative nature of bound-state problems, and the
complex structure of the QCD vacuum, suggests that one should employ dressed
gluon and quark propagators in constructing the kernel of the BSE. These can
be obtained via Dyson-Schwinger equations [DSEs].  The Bethe-Salpeter
equation is one type of DSE.

There have been many studies of meson and diquark spectroscopy using this
framework; a summary can be found in Ref.~\cite{DSErev}.  Typically, such
studies employ an Ansatz for the gluon propagator in solving the
rainbow-approximation quark-DSE and pair the input gluon propagator and
calculated quark propagator to construct the dressed-ladder approximation
kernel for the meson/di\-quark BSE.  The BSE is then solved to obtain the
spectrum.  This rainbow-ladder truncation has the feature that Goldstone's
theorem is manifest; i.e., in the chiral limit, when the current-quark mass
$m_q=0$, the pion is a zero-mass bound-state of a strongly-dressed quark and
antiquark~\cite{DS79}.  With a few-parameter model for the gluon propagator,
it can be used to provide a good description of the light-light, light-heavy
and heavy-heavy meson spectrum and decays~\cite{JM93}.  However, it has the
defect that it admits $(\bar 3)_c$-diquark bound-states~\cite{PCR89}; such
coloured states have not been observed.

Herein we consider a truncation scheme that allows for a systematic
improvement in the construction of the kernels of the quark-DSE and
meson/diquark-BSE.  This procedure ensures that the pion remains a Goldstone
boson at every order.  The first correction to ladder-rainbow approximation
introduces a repulsive term in the Bethe-Salpeter kernel.  Using a model
gluon propagator, singular in the infrared and with no Lehmann
representation, we solve the quark DSE and obtain a quark propagator that
also has no Lehmann representation.  Pairing these gluon and quark
propagators in the meson channel, the repulsive term is almost cancelled by
attractive terms of the same order and therefore the higher order terms only
lead to a small change in meson masses.  However, due to the algebra of
$SU(3)_c$, the repulsive term is stronger in the diquark channel and is not
cancelled by the attractive terms.  This entails that there is no stable
diquark bound-state.

{\bf 2. Dyson-Schwinger and Bethe-Salpeter equations}.\\ 
%\section{Dyson-Schwinger and Bethe-Salpeter equations}
In Euclidean metric, with $\gamma_\mu$ hermitian and
$\{\gamma_\mu,\gamma_\nu\}=2\delta_{\mu\nu}$, the propagator for a dressed
quark has the form \mbox{$S(p) = -i\gamma\cdot p\,\sigma_V(p^2) +
\sigma_S(p^2)$}.  It is obtained as a solution of the quark DSE
\begin{eqnarray}
\label{fullDSE}
\lefteqn{ S^{-1}(p) \equiv i \gamma\cdot p + m_q + \Sigma(p)}\\
& & \nonumber
= i \gamma\cdot p + m_q + g^2 \int
     \sfrac{d^4k}{(2\pi)^4}\, \gamma_\mu \frac{\lambda^a}{2}\,
S(k)\, \Gamma_\nu^g (k,p) \frac{\lambda^a}{2}\,D_{\mu \nu}(p-k), 
\end{eqnarray}
where $m_q$ is the current-quark mass, $\Gamma_\mu^g(k,p)$ is the dressed
quark-gluon vertex and $D_{\mu\nu}(k)$ is the dressed gluon propagator,
which, in Landau gauge, can be written
\begin{equation}
\label{generalD}
g^2\,D_{\mu\nu}(k) \equiv 
\left(\delta_{\mu\nu} - \frac{k_\mu k_\nu}{k^2}\right)\,\Delta(k^2)~.
\end{equation}

The mass and Bethe-Salpeter amplitude for a quark-antiquark bound-state,
$\Gamma_M$, is obtained by solving
\begin{eqnarray}
\label{bsemeson}
\lefteqn{\Gamma^{EF}_M(p;P) = }\\
& & \nonumber
\int\,\sfrac{d^4k}{(2\pi)^4} \,K_M^{EF;GH}(k,p;P)\,
\left(S(k+\sfrac{1}{2}P)\Gamma_M(k;P)S(k-\sfrac{1}{2}P)\right)^{GH}~,
\end{eqnarray}
where $K_M(k,p;P)$ is the kernel, $P$ is the total-momentum of the system,
$k,p$ are the internal and external relative quark-antiquark momenta and the
superscripts are associated with the colour, flavour and Dirac structure of
the amplitude; i.e., $E=\{i_c,i_f,i_D\}$.

The analogue of Eq.~(\ref{bsemeson}) for quark-quark systems is
\begin{eqnarray}
\label{bsediquark}
\lefteqn{\Gamma^{EF}_D(p;P) = }\\
&&\nonumber
\int\,\sfrac{d^4k}{(2\pi)^4} \,K_D^{EF;GH}(k,p;P)\,
\left(S(k+\sfrac{1}{2}P)\Gamma_D(k;P)S^T(-k+\sfrac{1}{2}P)\right)^{GH}~,
\end{eqnarray}
where ``$T$'' denotes matrix-transpose.  The absence of a solution to this
equation entails that diquarks do not appear in the strong-interaction
spectrum.

In the isovector channel, each contribution to $K_M^{EF;GH}(k,p;P)$ has a
direct analogue in $K_D^{EF;GH}(k,p;P)$; an analogue that can be obtained via
the replacement
\begin{equation}
\label{mesontodiquark}
S(k)\,\gamma_\mu\,\frac{\lambda^a}{2} \to 
\left[\gamma_\mu\,\frac{\lambda^a}{2}\,S(-k)\right]^T~,
\end{equation}
($\{\lambda^a/2\}_{a=1}^8$ are the generators of $SU(3)_c$) in each antiquark
segment of the meson kernel, which can be traced unambiguously from the
external antiquark-line of the meson amplitude.  This means that there are no
contributions in the diquark channel that are not also present in the
isovector meson channel.

{\it Rainbow-ladder approximation}.\\ The rainbow approximation to the quark
DSE is defined by the choice $ \Gamma_\mu^g(k,p) \equiv \gamma_\mu $ in
Eq.~(\ref{fullDSE}); the ladder-approximation to the meson BSE by
\begin{eqnarray}
\lefteqn{K_M^{EF;GH}(k,p;P)\,
\left(S(k+\sfrac{1}{2}P)\Gamma_M(k;P)S(k-\sfrac{1}{2}P)\right)^{GH}}\\
&& \nonumber
\equiv -g^2\,D_{\mu\nu}(p-k)\,
\left(\gamma_\mu\,\frac{\lambda^a}{2}\,
S(k+\sfrac{1}{2}P)\Gamma_M(k;P)S(k-\sfrac{1}{2}P)
\gamma_\nu\,\frac{\lambda^a}{2}\,\right)^{EF}
\end{eqnarray}
in Eq.~(\ref{bsemeson}).  
%--
% For the diquark,
% \begin{eqnarray}
% \lefteqn{K_D^{EF;GH}(k,p;P)\,
% \left(S(k+\sfrac{1}{2}P)\Gamma_D(k;P)S^T(-k+\sfrac{1}{2}P)\right)^{GH}}\\
% && \nonumber
% \equiv g^2\,D_{\mu\nu}(p-k)\,
% \left(\gamma_\mu\,\frac{\lambda^a}{2}\,
% S(k+\sfrac{1}{2}P)\Gamma_D(k;P)\left[\gamma_\nu\,\frac{\lambda^a}{2}\,
% S(-k+\sfrac{1}{2}P)\right]^T\,\right)^{EF}
% \end{eqnarray}
% in Eq.~(\ref{bsediquark}), which illustrates the application of
% Eq.~(\ref{mesontodiquark}).  
%--

In the chiral limit the rainbow-approximation DSE and the pseudoscalar-meson
BSE in ladder-approximation are equivalent when $P^2=0$; i.e., one
necessarily has a massless, pseudoscalar bound-state when chiral symmetry is
dynamically broken\cite{DS79}.  Goldstone's theorem is manifest.  In any
truncation, such an identity between the quark DSE and the BSE in the
isovector, pseudoscalar, meson channel is sufficient to ensure that this
meson is a Goldstone boson~\cite{HJM95}.  This provides for a straightforward
understanding of the dichotomy of the pion as both a Goldstone boson and
quark-antiquark bound-state.

A gluon propagator that has no Lehmann representation may be interpreted as
describing a confined particle.  Such a gluon propagator can, via
Eq.~(\ref{fullDSE}), yield a quark propagator with the same
property~\cite{MN83}.  Hence, in rainbow-ladder approximation, one can
develop a spectroscopic model of mesons with confined quarks and gluons.  

However, at this order the diquark BSE, Eq.~(\ref{bsediquark}), admits
solutions; i.e., one has $(\bar 3)_c$ diquark states in the
spectrum~\cite{PCR89}, which are not observed.  (It follows from the algebra
of the generators of $SU(3)_c$ that, in this approximation, there are no
$(6)_c$ diquark solutions just as there are no $(8)_c$ meson solutions.)

{\bf 3. Next-order Truncation}\\
%\section{Next-order Truncation}
As the next level of truncation for the quark DSE we consider
Eq.~(\ref{fullDSE}) with
\begin{eqnarray}
\label{vertexcorrect}
\Gamma_\nu^g(k,p) =
\gamma_\nu + \frac{1}{6}\int\frac{d^4 l}{(2\pi)^4}\,
g^2 D_{\rho\sigma}(p-l)\,
\gamma_\rho\,S(l+k-p)\gamma_\nu\,S(l)\gamma_\sigma~.
\end{eqnarray}
This is the first-order correction of the vertex by the dressed-gluon
propagator and the dressed-quark propagator, which is obtained as the
solution of this DSE.  Here we omit the explicit $3-$gluon vertex that could
contribute at this order.  This entails that we explore $3-$ and $4-$body forces
only to the extent that they are incorporated via the nonperturbative
dressing of the gluon propagator.

For the purpose of illustration and clarity we employ a model gluon
propagator~\cite{MN83} that allows for an algebraic solution of the
Dyson-Schwinger and Bethe-Salpeter equations:
\begin{equation}
\label{modelD}
\Delta(k^2) \equiv 16\pi^4\,G\,\delta^4(k)~,
\end{equation}
in Eq.~(\ref{generalD}), where $G=\eta^2/4$ with $\eta$ a mass scale.
This one-parameter model possesses the infrared enhancement, due to the
$3-$gluon vertex, suggested by the studies of Refs.~\cite{IRBig} but
underestimates the interaction strength for $k^2>0$.  This is appropriate in
a study whose focus is confinement and dynamical chiral symmetry breaking,
which are infrared effects.  The qualitative features of our results are not
sensitive to this choice.

Using Eqs.~(\ref{vertexcorrect}) and (\ref{modelD}) in Eq.~(\ref{fullDSE})
one obtains\footnote{In deriving this equation we have used the
identity\\ \mbox{$\displaystyle \int d^4
l\,\delta^4(l)\,\frac{p\cdot l k\cdot l}{l^2}\, f(k,p,l) \equiv \sfrac{1}{4}
\int d^4 l\,\delta^4(l)\,p\cdot k\,f(k,p,l)$}}
\begin{equation}
\label{newDSE}
 S^{-1}(p) = i \gamma\cdot p + m_q + 
G\,\gamma_\mu S(p) \gamma_\mu + \frac{1}{8}\, G^2\, 
\gamma_\mu S(p) \gamma_\nu S(p) \gamma_\mu S(p)\gamma_\nu~.
\end{equation}
Neglecting the O$(G^2)$ term yields the rainbow-approximation quark DSE.

\begin{figure}[h,t]
  \centering{\
     \epsfig{figure=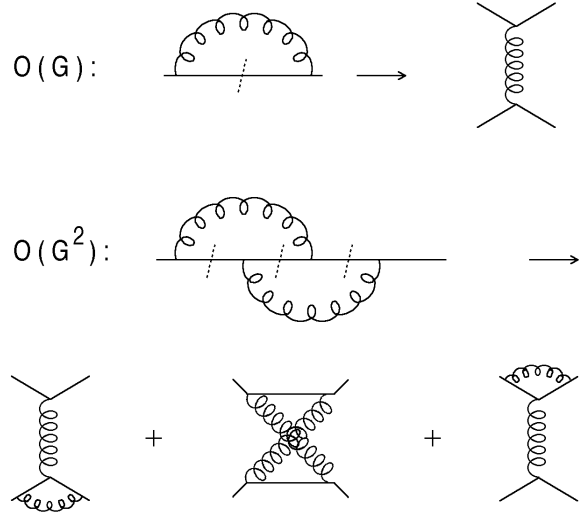,height=7cm,rheight=7cm}  }
\caption{\label{OGtwo} The replacement procedure of
Eq.~(\protect\ref{DSEtoBSE}) illustrated to O$(G^2)$, which provides the
kernel of the meson and diquark Bethe-Salpeter equations.  The internal solid
lines represent dressed quark and gluon propagators.}
\end{figure}

{\it Goldstone Theorem}.\\ 
An important consideration in extending the kernel of the Bethe-Salpeter
equation is to ensure that in doing so one preserves the Goldstone boson
character of the pion.  To this end we observe that the ladder-approximation
kernel in the meson BSE can be obtained directly from the expression for
$\Sigma(p)$ in the rainbow-approximation quark DSE via the replacement
\begin{equation}
\label{DSEtoBSE}
\gamma_\mu\,S(k)\,\gamma_\nu \to 
\gamma_\mu\,S(k+P/2)\,\Gamma_M(k,P)\,S(k-P/2)\,\gamma_\nu~,
\end{equation}
which is illustrated in the top diagram of Fig.~\ref{OGtwo}.  

In the isovector channel, this procedure can be implemented at every order;
i.e., in every term of the quark DSE one may sequentially replace each
explicit, internal quark propagator in this way, which is illustrated for
Eq.~(\ref{newDSE}) in Fig.~\ref{OGtwo}.  This generates all contributions of
a given order to the kernel and ensures that Goldstone's theorem is preserved
at that order, as will become clear below.  Applying the procedure to vacuum
polarisation insertions does not generate additional terms in the isovector
kernel; a fact that very much simplifies the study of isovector systems,
allowing one to employ a model gluon propagator and maintain Goldstone's
theorem.
%%--------------------
%%--------------------

These arguments are independent of the form of the model gluon propagator.
The inclusion of $3-$ and $4-$gluon vertices is a straightforward extension,
which does not modify the conclusions.

The isoscalar channel receives additional contributions from vacuum
polarisation diagrams.  This may provide the basis for understanding the
$\eta$-$\eta$' mass splitting in the present framework\cite{etaetap}.

The full O$(G^2)$ kernel for the isovector-meson Bethe-Salpeter equation is
illustrated in Fig.~\ref{OGtwo}.  Using Eq.~(\ref{modelD}), the
Bethe-Salpeter equation for $(1)_c$, isovector mesons is
\begin{eqnarray}
\label{mesonOGG}
\lefteqn{\Gamma_M(p;P) = 
- G\,\gamma_\mu\,\chi_M\,\gamma_\mu}\\
&& \nonumber
- \frac{1}{8}\,G^2\,\gamma_\mu\,\left(
S_+\,\gamma_\nu\,S_+\,\gamma_\mu\,\chi_M
+ \underline{S_+\,\gamma_\nu\,\chi_M\,\gamma_\mu\,S_-}
+ \chi_M\,\gamma_\nu\,S_-\,\gamma_\mu\,S_-
\right)\,\gamma_\nu
\end{eqnarray}
where $S_\pm \equiv S(p\pm P/2)$ and $\chi_M\equiv S_+\,\Gamma_M(p;P)\,S_-$.
Using Eq.~(\ref{mesontodiquark}) the Bethe-Salpeter equation for $(\bar
3)_c$, isovector diquarks can be written
\begin{eqnarray}
\label{diquarkOGG}
\lefteqn{\Gamma_{D^{\bar 3}}^C(p;P) = 
- \frac{1}{2}G\,\chi_{D^{\bar 3}}^C\,\gamma_\mu
- \frac{1}{16}\,G^2\,\gamma_\mu\,\left(
S_+\,\gamma_\nu\,S_+\,\gamma_\mu\,\chi_{D^{\bar 3}}^C
\right.}\\
&& \nonumber
\left.
+ 5\,\underline{
S_+\,\gamma_\nu\,\chi_{D^{\bar 3}}^C\,\gamma_\mu\,S_-}
+ \chi_{D^{\bar 3}}^C\,\gamma_\nu\,S_-\,\gamma_\mu\,S_-
\right)\,\gamma_\nu~,
\end{eqnarray}
with $\Gamma_{D^{\bar 3}}\equiv\Gamma_{D^{\bar 3}}^C C$ and
$\chi_{D^{\bar3}}^C\equiv S_+\,\Gamma_{D^{\bar 3}}^C(p;P)\,S_-$, where
$C=\gamma_2\gamma_4$ is the charge conjugation matrix.  In these equations
the underlined term is the ``crossed-box'' contribution of Fig.~\ref{OGtwo}.

{\it Diquark bound-states}.\\
At O$(G)$ the only difference between Eqs.~(\ref{mesonOGG}) and
(\ref{diquarkOGG}) is a 50\% reduction in the coupling strength; i.e., the
coupling is twice as strong in the meson equation.  This follows from the
algebra of the $SU(3)_c$ generators and entails~\cite{CRP89} the existence of
scalar and pseudovector diquark bound-states with $m_{qq}^{\rm 0^+}>m_{\bar q
q}^{0^-}$ and $m_{qq}^{\rm 1^+}>m_{\bar q q}^{1^-}$.

Of the terms in Eqs.~(\ref{mesonOGG}) and (\ref{diquarkOGG}) only the
underlined one in each is repulsive in the $0^-_{\bar q q}$, $1^-_{\bar q
q}$, $0^+_{q q}$, $1^+_{q q}$ channels.  The numerical factors arise from the
algebra of the $SU(3)_c$ generators and do not depend on the form of the
gluon propagator.  Relative to its companion, parenthesised contributions,
this term is 5-times as large in the diquark equation, which provides for the
possibility that it eliminates diquark bound-states.

\begin{figure}[h,t]
  \centering{\
     \epsfig{figure=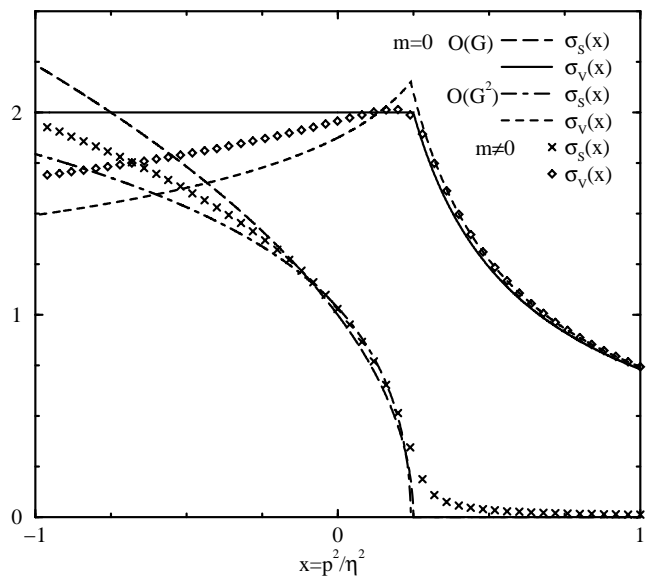,height=9cm,rheight=9cm}  }
\caption{\label{DSEsoln} We plot $\bar\sigma_V(x)$ and $\bar\sigma_S(x)$
obtained as the solution of Eq.~(\protect\ref{newDSE}); $\sigma_V(p^2)=
\bar\sigma_V(x)/\eta^2$, $\sigma_S(p^2)= \bar\sigma_S(x)/\eta$.}
\end{figure}

{\bf 4. Results and Conclusions}\\
%\section{Results and Conclusions}
In order to study the bound-state spectrum one must solve Eq.~(\ref{newDSE}).
Our results are illustrated in Fig.~\ref{DSEsoln}.  The O$(G^2)$ corrections
become noticeable for $p^2\lsim \eta^2/2$, which is important because it is the
domain sampled in the Bethe-Salpeter equation.  It is also important to note
that one has dynamical chiral symmetry breaking for any $G>0$; i.e., for
$m_q=0$: $\sigma_S(0)>0$, $\forall G>0$.  Further, $\sigma_V$ and $\sigma_S$
have no pole on the real-$p^2$ axis.  Therefore the quark propagator has no
Lehmann representation and can be interpreted as describing a confined
particle.

The model gluon propagator of Eq.~(\ref{modelD}) entails that the bound-state
constituents have zero relative momentum; i.e., $p=0$.  In this case the most
general form of the $0^-_{\bar q q}$ Bethe-Salpeter amplitude is
\begin{equation}
\label{pibsaform}
\Gamma_{M}^{0^-}(P)= 
\left[\theta_1^{0^-}(P^2) 
+ i\frac{\gamma\cdot P}{\eta} \theta_2^{0^-}(P^2)\right]\gamma_5~.
\end{equation}
Substituting Eq.~(\ref{pibsaform}) into Eq.~(\ref{mesonOGG}), with $p=0$, one
obtains a $2\times 2$ matrix eigen-value problem of the form \mbox{$H
\,\Theta = \Theta$}; where the elements of $H$ depend on $P^2$, and
$\Theta^T=(\theta_1,\theta_2$).  The eigen-value problem is solved when that
$P^2$ is found for which \mbox{$det\{ H(P^2)-I\} =0$}; the eigen-vector
follows immediately.  The same procedure is followed in each of the other
channels.  The vector-meson Bethe-Salpeter amplitude has the form
\begin{eqnarray}
\label{rhobsaform}
\Gamma_{M}^{1^-}(P)& = &\epsilon^\lambda\cdot\gamma \theta_1^{1^-}(P^2) 
+ \frac{i}{\eta}\sigma_{\mu\nu}\epsilon_\mu^\lambda P_\nu
\theta_2^{1^-}(P^2)~,
\end{eqnarray}
with $\epsilon_\mu^\lambda(P)$, $\lambda=0,\,\pm 1$, the polarisation vector:
$\epsilon^\lambda\cdot P=0$; $\Gamma_{D^{\bar 3}}^{0^+\,C}$ has the same form
as the pseudoscalar meson amplitude in Eq.~(\ref{pibsaform}); and
$\Gamma_{D^{\bar 3}}^{1^+\,C}$ the same form as the vector-meson amplitude in
Eq.~(\ref{rhobsaform}).

Our chiral limit results are presented in
Table~\ref{masstable}.  The O$(G)$:O$(G)$ eigen-vectors are 
$(\theta_1^{0^-},\theta_2^{0^-})=(0.83,0.55)$, 
$(\theta_1^{1^-},\theta_2^{1^-})=(1,0)$.  
At O$(G^2)$:O$(G^2)$ they are 
$(\theta_1^{0^-},\theta_2^{0^-})=(0.87,0.49)$, 
$(\theta_1^{1^-},\theta_2^{1^-})=(0.99,-0.12)$.

% ; the eigen-vectors in Eq.~(\ref{mesoneigen}).
%--
\begin{table}[h,t]
\begin{center}
\caption{\label{masstable} Calculated meson and diquark masses, in GeV
($\eta=1.06$~GeV). The labels ``O$(G^n)$: O$(G^m)$'' mean that the solution
of the O$(G^n)$ quark Dyson-Schwinger equation was used in the O$(G^m)$
Bethe-Salpeter equation.  The non-zero quark mass was chosen to reproduce the
experimental ratio $m_\pi/m_\rho$ at O$(G^2)$:O$(G^2)$.  ``Unbound'' means
that there is no solution of the associated homogeneous Bethe-Salpeter
equation for real $P^2$.}
\begin{tabular}{llll}
$m_q=0$                   & O$(G)$: O$(G)$ 
                                & O$(G)$: O$(G^2)$
                                                & O$(G^2)$: O$(G^2)$ \\ \hline
$m_{\bar q q}^{0^-}$    & 0     & 0.30          & 0                     \\
$m_{qq}^{0^+} $         & 1.19  & Unbound       & Unbound               \\
$m_{\bar q q}^{1^-}$    & 0.750 & 0.745         & 0.823                 \\
$m_{qq}^{1^+} $         & 1.30  & Unbound       & Unbound               \\\hline
$m_q=0.012$             &       &               &                    \\ \hline
$m_{\bar q q}^{0^-}$    & 0.140 & 0.328         & 0.136                 \\
$m_{qq}^{0^+} $         & 1.21  & Unbound       & Unbound               \\
$m_{\bar q q}^{1^-}$    & 0.767 & 0.760         & 0.770                 \\
$m_{qq}^{1^+} $         & 1.32  & Unbound       & Unbound               \\\hline
\end{tabular}
\end{center}
\end{table}
%--
% \begin{equation}
% \label{mesoneigen}
% \begin{array}{lclcl}
%  & & {\rm O}(G){\rm :O}(G) & & {\rm O}(G^2){\rm :O}(G^2) \\
% (\theta_1^{0^-},\theta_2^{0^-})& &  (0.83,0.55) && (0.87,0.49) \\
% (\theta_1^{1^-},\theta_2^{1^-}) & & (1,0) & & (0.99,-0.12)
% \end{array}
% \end{equation}
%--

The Goldstone theorem is manifest when the quark DSE and pseudoscalar meson
BSE are truncated consistently; i.e, at O$(G)$:O$(G)$ and O$(G^2)$:O$(G^2)$.
This can be shown analytically. If the dressing is inconsistent; e.g.,
O$(G)$:O$(G^2)$, the pseudoscalar is half as massive as the vector meson.

The O$(G^2)$ corrections only provide for a small (10\%) mass increase in the
vector-meson channel, as one would expect of a weak, net-repulsive
interaction.  It is weak because of the cancellation between the
vertex-correction and crossed-box contributions, which is a necessary
consequence of the preservation of Goldstone's theorem.

In the diquark channel, however, where the coefficient of the repulsive term
is larger and the cancellation incomplete, the O$(G^2)$ corrections have the
significant effect of eliminating the bound-state pole on the real-$P^2$
axis.

We note that the pseudovector, $\theta_2^{0^-}$, component of the
pseudoscalar meson is a significant part of its Bethe-Salpeter amplitude
whereas the tensor, $\theta_2^{1^-}$, component of the vector meson is small.
This feature is also observed in the separable Ansatz studies of
Ref.~\cite{bsesep}.

Our results for $m_q \neq 0$ are presented in Table~\ref{masstable}.  The
O$(G^2)$:O$(G^2)$ meson eigen-vectors are:
\mbox{$(\theta_1^{0^-},\theta_2^{0^-})= (0.86,0.51)$} and
\mbox{$(\theta_1^{1^-},\theta_2^{1^-}) = (0.99,-0.13)$}.
%-- 
% in Eq.~(\ref{mesoneigenm}).
% \begin{equation}
% \label{mesoneigenm}
% \begin{array}{lclcl}
% (\theta_1^{0^-},\theta_2^{0^-})&& (0.86,0.51) \\
% (\theta_1^{1^-},\theta_2^{1^-}) & & (0.99,-0.13)
% \end{array}
% \end{equation}
%-- 
The Goldstone boson character of the pseudoscalar is clear; i.e., at a
consistent level of truncation its mass increases rapidly from zero as $m_q$
is increased.  In contrast, the vector mass shifts upward by $<$~1\%. (The
upward shift is $\approx$~10\% for $m_q=0$ because the difference between the
O$(G)$ and O$(G^2)$ solutions of the quark DSE is greater in this case.)  A
non-zero current-quark mass introduces little quantitative and no qualitative
change in the diquark channels; i.e., they remain unbound.

\begin{figure}[h,t]
  \centering{\
     \epsfig{figure=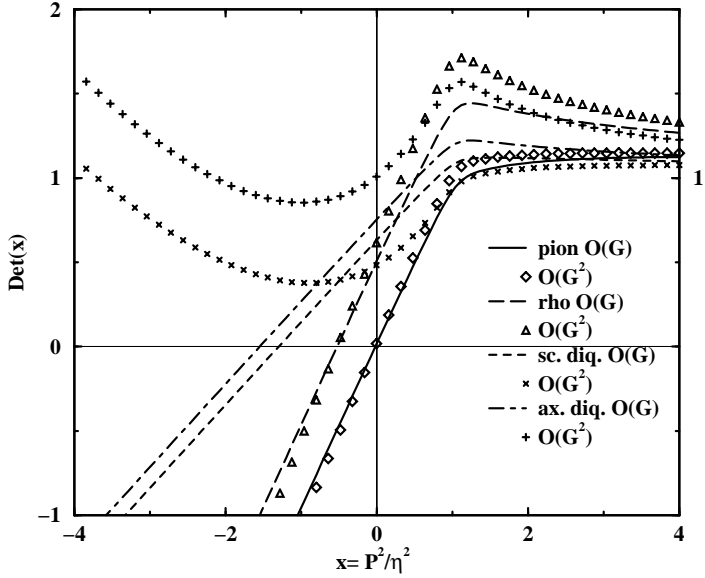,height=9cm,rheight=9cm}  }
\caption{\label{detHmI}  $det\{H(P^2)-I\}$ plotted as a function of $P^2$.  This
function vanishes at the square of the bound-state mass in the channel under
consideration.}
\end{figure}

For $m_q=0.012$~GeV we have plotted $det\{H(P^2)-I\}$ in Fig.~\ref{detHmI}.
It illustrates the effect of the O$(G^2)$ repulsive term in the
Bethe-Salpeter kernel, which shifts the zero in the meson channel very little
but completely eliminates it in the diquark channel.  The additional
repulsive strength in the diquark channel is amplified by the growth in
$\sigma_S(x)$ for $x<0$; i.e., the feature of confinement manifest in the
quark propagator, and due to the infrared enhancement of the gluon
propagator, plays a role in destabilising the diquarks.  
% The $3-$gluon vertex is responsible for this enhancement of the gluon
% propagator in the infrared~\cite{IRBig}.
%--
%--

A motivation for our study is the observation that while the absence of a
Lehmann representation for gluon and quark propagators ensures there are no
gluons and quarks in the strong interaction spectrum, it neither entails nor
precludes the existence of bound-states, whether coloured or not.  We have
shown that the rainbow-ladder truncation is peculiar in the sense that it
alone has no repulsive terms in the BSE kernel.  In every truncation beyond
this there are both attractive and repulsive terms.  We saw that in a
consistent truncation there was almost complete cancellation between the
O$(G^2)$ attractive and repulsive terms in the meson channel.  This is an
indication of why the studies of meson spectroscopy and decays using a model
gluon propagator in rainbow-ladder approximation have been successful.  In
the diquark channel the situation is quite different.  The algebra of
$SU(3)_c$ ensures that the coefficient of the O$(G^2)$ repulsive term is
larger and this, coupled with confinement as manifest in the form of the
quark propagator, entails that the repulsive effect survives to ensure the
absence of a stable diquark bound-state.  This effect cannot reasonably be
reproduced in rainbow-ladder approximation.  It is plausible that these
features persist at higher order and with the inclusion of explicit $3-$ and
$4-$gluon vertices.

%\parbox{138mm}{
{\bf Acknowledgments}.\\ This work was supported by the US Department of
Energy, Nuclear Physics Division, under contract number W-31-109-ENG-38. The
calculations described herein were carried out using the resources of the
National Energy Research Supercomputer Center.
%}

%______________________________ References ______________________________

%___________________________________________________________________________
\end{document}